\documentclass[sigconf,screen]{acmart}

\usepackage{amsmath}
\usepackage{multirow}
\usepackage{xcolor}
\usepackage{array}
\usepackage[flushleft]{threeparttable}

\newcommand{\beginsupplement}{%
        \setcounter{table}{0}
        \renewcommand{\thetable}{S\arabic{table}}%
        \setcounter{figure}{0}
        \renewcommand{\thefigure}{S\arabic{figure}}%
     }

\AtBeginDocument{%
  \providecommand\BibTeX{{%
    \normalfont B\kern-0.5em{\scshape i\kern-0.25em b}\kern-0.8em\TeX}}}

\copyrightyear{2022}
\acmYear{2022}
\setcopyright{rightsretained}
\acmConference[AIES'22]{Proceedings of the 2022 AAAI/ACM Conference
on AI, Ethics, and Society}{August 1--3, 2022}{Oxford, United Kingdom}
\acmBooktitle{Proceedings of the 2022 AAAI/ACM Conference on AI,
Ethics, and Society (AIES'22), August 1--3, 2022, Oxford, United
Kingdom}\acmDOI{10.1145/3514094.3534200}
\acmISBN{978-1-4503-9247-1/22/08}

\acmSubmissionID{aies227}

\begin{document}

\title{Transparency, Governance and Regulation of Algorithmic Tools Deployed in the Criminal Justice System: a UK Case Study}


\author{Miri Zilka}
\orcid{0000-0001-9640-8139}
 \affiliation{%
   \institution{University of Cambridge}
   \city{Cambridge}
   \country{UK}}
\affiliation{%
  \institution{The Alan Turing Institute}
  \city{London}
  \country{UK}}
\authornote{Both authors contributed equally to this research.}
 \email{mz477@cam.ac.uk}

\author{Holli Sargeant}
\orcid{0000-0003-3482-7789}
 \affiliation{%
   \institution{University of Cambridge}
   \city{Cambridge}
   \country{UK}}
\authornotemark[1]
 \email{hs775@cam.ac.uk}

\author{Adrian Weller}
\orcid{0000-0003-1915-7158}
 \affiliation{%
   \institution{University of Cambridge}
   \city{Cambridge}
   \country{UK}}
\affiliation{%
  \institution{The Alan Turing Institute}
  \city{London}
  \country{UK}}
\email{aw665@cam.ac.uk}

\renewcommand{\shortauthors}{Zilka et al.}

\begin{abstract}
We present a survey of tools used in the criminal justice system in the UK in three categories: data infrastructure, data analysis, and risk prediction. Many tools are currently in deployment, offering potential benefits, including improved efficiency and consistency. However, there are also important concerns. Transparent information about these tools, their purpose, how they are used, and by whom is difficult to obtain. Even when information is available, it is often insufficient to enable a satisfactory evaluation. More work is needed to establish governance mechanisms to ensure that tools are deployed in a transparent, safe and ethical way. We call for more engagement with stakeholders and greater documentation of the intended goal of a tool, how it will achieve this goal compared to other options, and how it will be monitored in deployment. We highlight additional points to consider when evaluating the trustworthiness of deployed tools and make concrete proposals for policy.
\end{abstract}

\ccsdesc[500]{Social and professional topics}
\ccsdesc[500]{Social and professional topics~Governmental regulations}
\ccsdesc[300]{Social and professional topics~Race and ethnicity}
\ccsdesc[300]{Security and privacy~Privacy protections}
\ccsdesc[300]{Computing methodologies~Artificial intelligence}
\ccsdesc[300]{Computing methodologies~Machine learning algorithms}
\ccsdesc[300]{Computing methodologies~Machine learning approaches}

\keywords{Criminal Justice, Trustworthy AI, Policy, Governance and regulation, Algorithms in deployment}

\maketitle

\section{Introduction}

Law enforcement 
faces intense scrutiny 
with important questions being raised about the extent to which algorithmic tools might explain instances of 
over-policing or other negative community interactions 
with the police \citep{sankin_2021_crime}. More and more algorithms are in use within criminal justice systems around the world, deployed by law enforcement and often designed by the private sector \citep{ensign2018runaway, mamalian1999use, weisburd2008compstat, babuta2020data, jansen2018data, sprick2019predictive}. There is a potential for promising benefits for law enforcement, including support for improved decision-making, efficiency, and consistency. However, there are many risks to their deployment. These systems have 
super-human abilities to analyse vast amounts of data which could be a benefit; however, they have significant drawbacks, such as a lack of contextual understanding or common-sense reasoning, leading to concerns raised in this paper. While prior work has mainly focused on the analysis of single tools, this paper aims to consider the deployment of these tools within the bigger picture of the criminal justice system. As a case study, we focus on algorithmic tools used in the criminal justice system in the UK. As we address key issues based on use cases in the UK, we reveal principles and considerations that will be relevant for law enforcement agencies across jurisdictions. 

Academics have been bringing examples of algorithmic policing to light in recent years from the US \citep{eubanks_automating_2018,pasquale_black_2019,selbst_disparate_2018,propublica} to Australia \citep{edward_santow_intelligence_2021,australian_human_rights_commission_human_2021}. Recent court cases have demonstrated the challenges of using algorithmic systems in the criminal justice process.\footnote{\textit{R (on the application of Edward BRIDGES) v The Chief Constable of South Wales Police} [2020] EWCA Civ 1058 (`\textit{Bridges v South Wales Police}'); \textit{State of Wisconsin v. Loomis} 881 N.W.2d 749 (2016); \textit{State of Kansas v. John Keith Walls}, (2017).} In the UK, academics, charity commissioned reports and freedom of information requests have identified some of these relevant algorithmic tools (cited within this paper). However, significant opacity remains along with confusion about what tools are available, which are in use and how they are deployed. This paper aims to offer a non-exhaustive survey of tools used in the criminal justice system in the UK in the following categories: data infrastructure, data analysis, and risk prediction.

Policy and law-makers are struggling to keep pace with the rapid changes in technology. In particular, 
current regulation does not capture the different risks present in deploying algorithmic tools at different stages of the criminal justice pipeline. In this work, first, we define specific technologies. Second, we survey and identify the use of these technologies at different stages in the UK criminal justice system. Then, we consider the implications for policy and regulation in the context of design principles for trustworthy algorithmic systems.   

\section{Definitions}
\label{sec:def}
In this work, we focus on specific data-driven technologies that are deployed throughout the criminal justice system in the UK, in the following categories:

\begin{enumerate}
    \item \textbf{Data infrastructure} refers to software and tools primarily used to record, store or organise data. The ability to filter and perform basic search queries on data is included in this category. More advanced options such as visualisations, statistical analyses, risk scores or predictions are included in the next categories. 
    \item \textbf{Data analysis} refers to software and tools primarily used to analyse data to create insights informing practitioners and stakeholders. This category includes custom data dashboards, visualisations and forms of statistical analysis that do not include machine learning, predictions or automated creation of risk scores. 
    \item \textbf{Risk prediction} refers to software and tools primarily used for predicting future risk based on analysis of past data. This category can be split into two subcategories:
    \begin{enumerate}
        \item Predicting future crime spatiotemporally includes software and tools primarily used to estimate the volume, location and time of criminal activity that has yet to occur. The predictive estimate is usually based on analysing spatial and temporal patterns in past data. 
        \item Predicting future individual crime risk includes software and tools, including manual tools, primarily used to estimate the risk of an individual offending in the future based on previous offending and/or other attributes of the individual. 
    \end{enumerate}
\end{enumerate}
 
Other technologies that may be of interest to the community, which we do not cover in this work, include facial recognition, gait identification, number plate recognition, speaker identification, speech identification, lip-reading technologies, gunshot detection algorithms and social media monitoring. While The broad principles discussed in this work apply to these technologies as well, in this work we do not elaborate on any of their specific challenges. 

\section{Algorithmic decision aids in the criminal justice pipeline: UK case study}

\begin{figure*}[t]
    \centering
    \includegraphics[width=\linewidth]{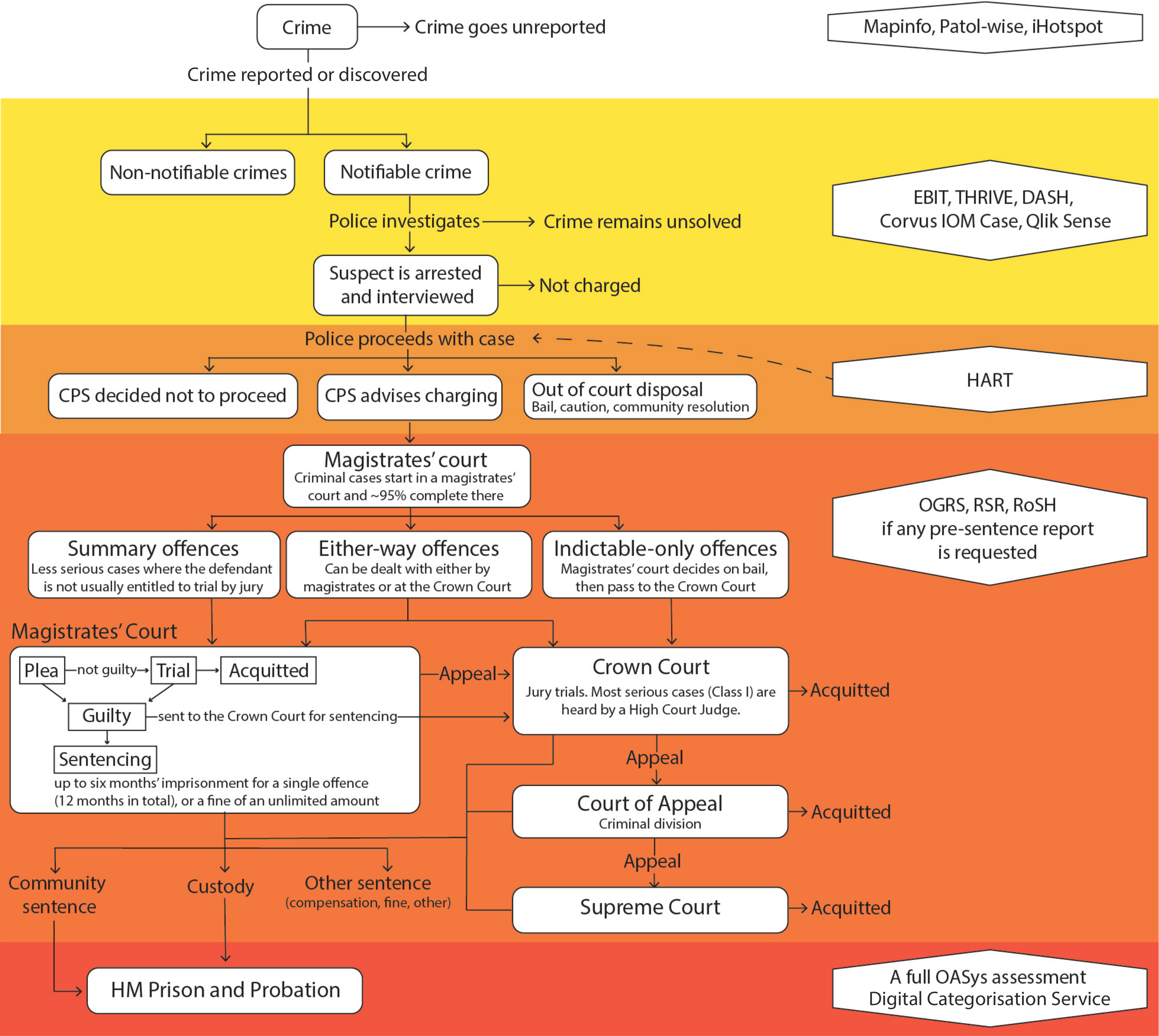}
    \caption{Stages in the criminal justice pipeline where predictive tools are used. In the top layer, tools are used to help decide where to look for crime. In all other layers, tools are used to help decide how to process an individual.}
    \label{fig:pipeline}
\end{figure*}

The criminal justice system in the UK has many players and components. Here we focus on four main components: law enforcement agencies, specifically the 43 police forces operating in England and Wales; the Crown Prosecution Service (CPS);  the courts, primarily referring to the magistrates’ court and the crown court; and Her Majesty's (HM) Prison and Probation Service. See Figure~\ref{fig:pipeline} for an illustration, to give context to the decisions that algorithmic systems are meant to assist. 

Only a portion of all illegal activity becomes known to law enforcement agencies. There are two main ways a criminal offence may become known: 
(i) A victim or a witness can report the crime to the police, or (ii) it can be discovered through proactive policing efforts, e.g., stop and search. The first decision point for the police after a crime is recorded is whether to dedicate resources to investigate the crime. Then, assuming an investigation was carried out and a suspect was arrested, the police need to decide whether to charge the suspect. The police can receive advice from the Crown Prosecution Service (CPS) and make the decision, or, more commonly, the case is transferred to the CPS, where a prosecutor decides if the case proceeds to court.  

Charges become convictions if the court finds the suspect guilty. In the UK, most offences ($\sim 95\%$) are heard in the Magistrates' Court. All offences start at the Magistrates' Court, where pre-trial custody status is decided. Serious criminal offences are tried in the Crown Court, where a judge presides over a jury trial. If the suspect is found guilty, a sentence will be decided by the magistrate or by a judge. In the UK, sentencing guidelines, setting minimum and maximum terms for the sentence, including limits on the time of imprisonment and the amount that can be set for monetary fines, must be followed unless ``it is contrary to the interests of justice to do so''. If the sentence includes an incarceration or probation period, the case is transferred to the HM Prison and Probation Service. In due course, the offender may be granted parole and released before serving their full sentence. 

This section includes representative examples of the tools and technologies described in Section~\ref{sec:def}, and examines how they are used across the criminal justice pipeline. To the best of the authors' knowledge, there is no comprehensive list of the technologies deployed within criminal justice in the UK. This work is not a complete list of such tools either -- there are many more tools in use even just by the police forces. However, the authors believe these examples are representative of the types of tools in use and the context of their deployment, in the UK, and likely beyond. The information we present is curated from government websites, published reports, peer-reviewed articles, news articles, freedom of information requests and conversations with police officers. We provide a reference for each tool. All tools we list have been in use, though some may no longer be in active use. 

Figure~\ref{fig:pipeline} illustrates where these tools are deployed in the criminal justice pipeline.

\subsection{Law Enforcement Agencies}

\subsubsection{Crime discovery and patrol planning}

\textit{Mapinfo} uses crime and incident data to map where crimes have occurred so police can monitor these `hotspots'. As this tool maps past crime, we classify it as a data analysis tool. West Midlands Police reported using this tool in answer to a freedom of information request from 2016 \citep{Law_Society}. \textit{Patrol-Wise} is an algorithm developed by University College London (UCL) in a research collaboration with West Yorkshire Police to predict burglaries on a street-by-street level \citep{Newsroom}. \textit{iHotSpot} is an AI-based predictive analytics engine used to predict daily crime incident hotspots. This tool was developed by SpaceTimeAI,  a spin-out company of UCL’s SpaceTimeLab. The system is integrated into the London Metropolitan Police as part of an EPSRC funded research collaboration \citep{iHotSpot, Cheng_2012}. Both \textit{Patrol-Wise} and \textit{iHotSpot} are predictive policing tools which aim to predict future crime spatiotemporally. We note the use of similar tools in the US, particularly \textit{Perdpol}, which was also trialled in the UK by Kent Police and the Metropolitan Police and was criticised for potential racial bias \citep{lum2016predict, sankin_2021_crime}. 

\subsubsection{Crime recording and investigation aids}
\label{sec:data_inf}
\paragraph{Data infrastructure.} 
The 43 police forces in England and Wales do not use a single primary data collection tool. For recording incidents as they are reported, and managing the first response, many forces use a platform called STORM Command and Control system \citep{STORM_Devon_and_Cornwall_Police,STORM_Dorset_Police,STORM_Essex_Police,STORM_Leicestershire_Police,STORM_Merseyside_Police,STORM_Staffordshire_Police,STORM_Suffolk_Police,STORM_Sussex_Police,STORM_West_Yorkshire_Police}.\footnote{STORM is a software product from Sopra Steria designed for emergency services, see https://www.soprasteria.co.uk/industries/public-safety.} 
If the reported incident is a crime or a recordable incident, it is also logged and managed in a record management system. Several of the forces use Niche RMS\footnote{Niche RMS is a private Canadian software company, see https://nicherms.com/.} for this purpose \citep{Niche,RUSI, STORM_Sussex_Police}. The data collected in relation to crime and reported incidents follows comprehensive national standards set by the Home Office.\footnote{The national standards for incidents and crime records set by the Home Office include The National Standard for Incident Recording (NSIR), National Crime Recording Standard (NCRS) and the Home Office Counting Rules (HOCR) \citep{STORM_Sussex_Police}.} To our knowledge, there are no national standards specifying which additional data fields may be collected by the forces, who each use their discretion. Information regarding which additional data fields are collected by each force is not immediately available. In some cases, information may exist within published police policies or can be accessed through freedom of information requests. The raw data collected by each force is not routinely accessible to other forces, but data sharing and data-based collaborations exist between forces. All forces make a series of annual data returns to the Home Office, containing aggregated data about crime, police workforce, arrests and stop and search \citep{homeoffice2021}. 


Police forces have access to several national datasets: 
The National Police Computer (NPC) stores and shares criminal record information across law enforcement agencies in the UK \citep{PNC}. The back-end data is not visible to individual forces. However, the forces routinely request information about an individual or against a vehicle registration number from the NPC. The forces can also request information from the Fingerprint and DNA national databases. These requests must be justified. 

\paragraph{Allocating investigative resources.} 

\textit{The Evidence-Based Investigative Tool (EBIT)} is a tool deployed by Kent Police to predict investigative success for minor, non-domestic assault and public order offences. A solvability assessment uses a logistic regression model, followed by a two-step review of the case. Using software built by Kent Police, the EBIT user answers eight questions and is provided with one of the following recommendations: further investigation, close the case pending further evidence, or further review by a supervisor \citep{mcfadzien2020evidence}.

\paragraph{Data analytics} 

\textit{The National Data Analytics Solution (NDAS)} is a project sponsored by the Home Office and led by West Midlands Police in collaboration with Accenture \citep{NDAS_Accenture}.\footnote{Accenture is a multinational company that provides consulting and professional services, see https://www.accenture.com/gb-en.} The project aims to develop ``algorithms on specific crime types that are applied to police-controlled data to produce insights and analysis which inform strategic and operational decision making'' \citep{Home_Office}. Currently, it is hosted by West Midlands Police and is used by seven additional police forces and the National Crime Agency.  

\textit{Sussex police} use customisable data analytic dashboards developed for them by a private company. For example, two dedicated dashboards were created in response to the Covid-19 pandemic: one to track employee self-isolation and one to track the daily changes in reported crimes, incidents, calls and arrests \citep{Sussex_Police}.

\subsubsection{Predicting risk associated with an individual.} 
Not all individual assessment tools involve the use of computational tools. Two manual risk assessment tools used nationally by the forces are \textit{THRIVE} and \textit{DASH}. Thrive stands for Threat, Harm, Risk, Investigation opportunities, Vulnerability of the victim and the Engagement level required to resolve the issue. It is used to assign a priority level to an incident \citep{THRIVE}. DASH stands for Domestic Abuse, Stalking and Harassment. It is a checklist tool used to assess the risk of cases involving domestic abuse, stalking, harassment and so-called honour-based violence \citep{DASH}.

\textit{The gangs violence matrix (GVM)} is a dataset of suspected London based gang members used by the Metropolitan Police. It is used as an intelligence tool ``to reduce gang-related violence, safeguard those exploited by gangs and prevent young lives from being lost'' \citep{matrix}. Each individual in the dataset is assigned a harm score and a victim score to indicate if they are likely to deliver or receive harm, respectively. Both scores are graded as Red, Amber or Green. Based on a critical report published by Amnesty International in 2018, the scores are automated risk scores, but no additional details are provided \citep{matrix,amnesty2018}. The report also highlighted concerns regarding stigma and bias in the database \citep{amnesty2018}. 

West Yorkshire Police uses an Integrated Offender Management (IOM) software system called Corvus IOM Case.\footnote{The tool is provided by Bluestar, a UK based company, see https://bluestar-software.co.uk/products/offender-management-iom/.} The system draws data from other sources, including STORM and Niche RMS (see Section~\ref{sec:data_inf}), analysing intelligence, crimes, arrests and substance misuse in order to derive an individualised score aimed at providing an indication of an individual’s likelihood to re-offend. Different crime types are scored differently, and the Risk of Re-Offending Cohort scores are categorically displayed as low, medium or high \citep{IOM}.

Avon and Somerset Police use a number of bespoke risk assessment tools. These tools are built within Qlik Sense, a data analytics platform developed by a private company.\footnote{Qlik Sense is developed by QlikTech International AB, see https://www.qlik.com/us/products/qlik-sense.} The tool produces an offender risk score in the range 0-100, which combines two predictions: the likelihood of offending combined with the potential harm if the re-offending does occur; and a vulnerability risk score to predict the likelihood of an individual becoming a victim of crime \citep{dencik2018}. 

\paragraph{Deciding whether to charge an individual.} 

Durham Constabulary uses an individualised \textit{Harm Assessment Risk Tool (HART)}. The tool was designed in collaboration with the University of Cambridge as part of the Checkpoint program to offer an alternative to prosecution for individuals classified as a medium risk -- likely to re-offend, but not in a serious violent manner. Qualifying individuals are offered a four month intervention program tailored to their needs and are not charged if they successfully complete it \citep{Oswald2018, Law_Society}.

\subsection{The Crown Prosecution Service and the Courts}

\paragraph{Data infrastructure.} A report on efficiency in the criminal justice system has highlighted that ``Problems occur as cases enter and progress through the system, including at crucial handover points between organisations, such as where information passes between police and the Crown Prosecution Service (CPS)'' \citep{Audit2016}. In an attempt to improve efficiencies, a single digital case management system was developed for the prosecution service and the courts, named the Common Platform, at an investment of hundreds of millions of pounds. The common platform was expected to be used in criminal hearings in all Crown and magistrates’ courts in England and Wales by the end of 2021 \citep{Common_Platform}.

\paragraph{Data analytics.} The Ministry of Justice has a ``largely centralised analytical function that provides wide-ranging analytical support to all of the Ministry’s activities'' \citep{MOJ}, including analysing data from the courts. The Ministry is taking steps to ensure the production of quality analysis with governance and assurance frameworks, open-source coding standards,\footnote{https://moj-analytical-services.github.io/our-coding-standards/}, and a Government Digital Service (GDS) data science cookie cutter.\footnote{https://github.com/ukgovdatascience/cookiecutter-data-science-gds} The latter is a standardized but flexible project structure for data science projects.

\subsubsection{Charging Decisions} 

Currently, the decision of whether to charge an individual with an offence within the Crown Prosecution Service (CPS) is done without algorithmic aids. The CPS plans to increase its use of AI, predominately to enable a more efficient search of text and video files, but there are no published plans to employ algorithms to predict risk or prosecution success \citep{CPS}.   

\subsubsection{Predicting future risk associated with an individual.} 
\label{sec:alg_court}
If a defendant pleads guilty or is found guilty during trial, the court can request a Pre-Sentence Report (PSR) to be prepared. The probation service writes the report after a private interview with the offender. Reports for magistrates’ court cases are usually same day PSRs, presented orally or in written format. Crown court cases may require a written report containing a full Offender Assessment System (OASys) assessment (see Section~\ref{OASys}), which requires the court to adjourn for 3 weeks. All reports include the following predictive scores: Offender Group Reconviction Score (OGRS); Risk of Serious Recidivism (RSR) Score; and a Risk of Serious Harm (RoSH) screening \citep{OASys2017}. 

\textit{Offender Group Reconviction Score (OGRS)} is a predictor of proven re-offending within 1 and 2 years of noncustodial sentence or discharge from custody. Proven re-offending only measures re-offending known to the authorities within the specified time frame and does not include arrests and offences pending prosecution. The OGRS4/G\footnote{OGRS4 is the fourth version of the tool. OGRS3 may still be in deployment.} includes all recordable offending and does not indicate the severity of predicted re-offending. Offenders with high OGRS4/G scores may re-offend rapidly but are comparatively unlikely to be involved in serious further offending. OGRS4/V includes violent proven re-offending only and outperforms OGRS4/G in predicting violent re-offending. OGRS prediction scores are based on a limited number of static risk factors\footnote{Static risk factors are factors that do not change or which change in only one direction. Examples include age and past criminal offences.} including age, gender and criminal history \citep{OASys2015}.

\textit{The Risk of Serious Recidivism (RSR)} was introduced in 2014 to predict the likelihood of an offender committing a seriously harmful re-offence within two years. The RSR tool provides three sub-scores, one for contact sexual re-offending, one for indecent image re-offending and one for non-sexual violence. All scores can be calculated from static variables, but non-sexual violence has an extended static and dynamic\footnote{Dynamic risk factors are factors about individuals or their environments that can change in either direction, e.g., employment, relationships, and substance misuse.} version that performs better than the brief static score \citep{OASys2015}.

\textit{Risk of Serious Harm (RoSH)} Serious harm is defined in this context as an ``event that is life-threatening and/or traumatic and from which recovery, whether physical or psychological, can be expected to be difficult or impossible'' \citep{OASys2015}. RoSH screening is done in all cases to indicate if a full analysis should be completed. The levels of RoSH are Low, Medium, High and Very High. Medium level indicates serious harm is unlikely unless there is a change in the offender’s circumstances. Very high indicates a high and imminent risk of serious harm. In the full RoSH analysis, there are separate scores indicating risk to children, known adults, the general public, staff and other prisoners. For all but the latter, there is a separate risk level for community and custodial settings \citep{OASys2015}.

\subsection{HM Prison and Probation Service}

\subsubsection{Security categorisation}
The \textit{digital categorisation service} (DCS) is an algorithmic tool used to make, record and justify prison security categorisation decisions. The tool ``highlights risk information'' and suggests an initial categorisation for the prisoner that is reviewed and can be changed by the staff \citep{Prison_Reform}. 

\subsubsection{Predicting future risk associated with an individual.}
\label{OASys}
The prison and probation services in England and Wales use the actuarial risk and needs assessment tool OASys. All offenders undergo a basic screening, and offenders with a sentence of 12 months or more undergo a full OASys assessment. OASys generates several risk scores based on static and dynamic risk factors to assess the likelihood of re-offending and risk of harm to self and others. The OGRS and RoSH scores mentioned in section 3.2.3 form part of the full OASys assessment \citep{oasys,Prison_Reform2018}. 

\section{Implications for policy and regulation}
The use of algorithmic systems -- whether computerised or manual -- raises the hope for significant benefits in terms of efficiency and consistency beyond those achievable by standard human methods. However, they also have significant drawbacks, such as a lack of contextual understanding or common-sense reasoning, leading to important concerns.

In the UK, algorithmic decision aids are in active deployment at almost every decision point in the criminal justice pipeline described above. The incorporation of new technologies is further increasing in the following years. The Home Office has indicated that they ``will therefore back and empower the police to use new technologies to deliver operational effect in a way that maintains public trust'' \citep{Home_Office2021}. Given the high-stakes nature of law enforcement decision-making, it is critical to design trustworthy algorithmic tools. In this section, we identify points to consider when evaluating the benefits and costs of deploying algorithmic tools compared to traditional analogue policing. We highlight key aspects required for evaluating the system as a whole, in addition to the individual tools. We suggest policy considerations for different stages in the data and criminal justice pipeline and in the context of different stakeholders.

\subsection{Design Principles for trustworthy algorithmic systems}
We consider the following principles of trustworthy use of tools in respect of different stages and stakeholders in the UK criminal justice system. Many of the points we mention have been discussed previously, e.g. see the ALGO-CARE framework for algorithms in policing \citep{Oswald2018} and a 10-point guide for evaluating if risk assessment tools are fit for purpose \citep{fazel2018}. 

\textbf{Fit for purpose:} A specific purpose for the collection of data and deployment of tools is necessary to justify whether the expected outcomes yield clear societal benefits. The appropriateness of its use must be assessed with respect to how it performs in deployment. The Metropolitan Police has stated that ``When policing uses technology, it does so to meet a defined policing purpose which must justify any privacy intrusion. The technology’s effectiveness and demographic performance (where relevant) must be assured as fit for its intended purpose, its use must be transparent to the degree possible within a policing context, and there must be safeguards, supported by community engagement and suitable oversight'' \citep{metpolice2021}. It is also important for the goal to be measurable. For example, Avon and Somerset Constabulary state they use ``advanced analytics along with data visualisation dashboards'' in order to ``support more effective decision making'' and improve ``understanding around demand, risk and threat'' \citep{Avon2021}. In this example, the aim is articulated; however, it is not clear if or how success is measured. We note that user satisfaction has been monitored and is reported to be high \citep{Avon2021}.  

\textbf{Data quality:} The quality of data and the design choices made at the collection stage can dramatically impact an algorithm's intended function. When in training and in use, data used in tools is often collected in a way that involves some variety of selection bias. 
For example, the digital categorisation tool used by the prison service in the UK relies on the quality of intelligence fed into it. However, it is not clear which, if any, data quality control procedures have been put in place. At the point of use, the intelligence cannot be easily verified or challenged. As a result, the automated categorisation produced by the tool cannot be easily verified or challenged either. Specific concerns have been raised around ``intelligence about BAME and Muslim men which cannot be effectively challenged, and which may reflect conscious or unconscious bias may be driving unfair treatment which the categorisation algorithm then cements in a way which will follow the prisoner throughout their custodial history" \citep{Prison_Reform}. 

\textbf{Fairness:} There is extensive literature reporting bias against minority groups within algorithms deployed in the criminal justice system \citep{selbst2017,chouldechova2017fair}. Many concerns focus on tools deployed in the US \citep{propublica}, but there is also significant worry about tools deployed in the UK \citep{amnesty2018,bigbrother2019}. A common concern is that an algorithm trained on biased data may learn to repeat or even reinforce the same bias unless steps are taken to avoid it. The baseline often used in this context is whether using the tool increases disproportionate outcomes. For example, a pilot study of a prison security categorisation tool didn't seem to suggest that using the new tool increased BAME disproportionately in higher-risk categories, but nor did it reduce it \citep{Prison_Reform}. We note even when disproportionately does not increase following the adoption of a tool, it does not necessarily mean the tool is unbiased. Avon and Somerset police highlight the exclusion of ethnicity as an input variable in their tools \citep{Avon2021}. However, not only does this not guarantee the tool is unbiased with respect to ethnicity but excluding the variable can sometimes aggravate the disproportionality in outcomes \citep{Joshua2021}. 
   
\textbf{Transparency and accountability:} For many reasons, it is difficult to ensure algorithms in criminal justice are fully transparent.\footnote{For further discussion of the challenges surrounding algorithmic transparency, see \citep{weller2019transparency,coyle2020explaining,bhatt2020machine}} It is important, however, that users and stakeholders have a sufficient understanding of the underlying mechanisms. It is key that accountability for decisions always remains with human decision-makers. Ensuring this requirement is met can prove more difficult when the tool is proprietary and is supplied by an external company, which is often the case. 

\textbf{Privacy:} The availability of data on many individuals enables useful patterns to be discovered by statistical or machine learning methods. However, from a legal and policy perspective, the protection of individual privacy is also important, and it can be hard to strike the right balance. It is important to protect individuals from invasive or disproportionate data collection, especially when the data is of a sensitive nature. This is of particular concern when data is shared between different agencies. For example, Avon and Somerset Police are sharing data with local agencies to ``collectively be better at identifying risk and vulnerability to support early intervention initiatives''. Specifically, they give an example that ``a model identifying a child at high risk of sexual exploitation would be validated through additional information such as referrals, social workers, analysts and other data held on systems before a practitioner
makes a decision'' \citep{Avon2021}. This example highlights the high level of causation required due to the level of sensitivity of the information and the vulnerability of the individuals in question. Avon and Somerset police state that ``Data Privacy Impact Assessments (DPIA)'' are filled for every tool \citep{Avon2021}. It is not clear if these assessments are publicly available and whether individuals can request to be excluded. 

\textbf{Robustness:} If an algorithmic system performs well in some settings, it can create the impression that it will perform well across all settings. However, algorithms can behave in fragile, hard to predict ways, such that small changes in input features can lead to unexpectedly large changes in model outputs \cite{szegedy2014intriguing}. Currently, in the UK, a new algorithmic tool can be introduced without a pilot or a trial period. In practice, many tools have been trialled by the different police forces and later abandoned. Unfortunately, the results and conclusions of these pilots are not often published.  

\textbf{Contestability:} The ability to contest decisions made or assisted by an automated system is an important safeguard in the trustworthy deployment of algorithmic tools \citep{mulligan2019shaping, ploug2020four, lyons2021fair}. 

Granting an individual recourse for decisions that they believe were not granted fairly requires a well-designed process. It may involve a formal internal review and a clear cause of action for judicial review. Such processes requires a level of transparency and explainability to ensure understanding of the algorithmic process and outcome.  

\subsection{Data infrastructure}
Data infrastructure and data collection standards are inconsistent across the 43 police forces in England and Wales, although there are some data sharing and database collaborations between forces. The Home Office has stated that work is underway to establish a Central Data Office within the Police Digital Service, which will ``provide the essential infrastructure for the sector'' \citep{homeoffice2021}. Where data infrastructure is not fit for purpose, organisations cannot benefit from the potential to improve criminal justice processes. Better data infrastructure would allow better information sharing between all 43 law enforcement agencies to the make performance of their duties easier and potentially improve public safety. The UK National Policing Digital Strategy recognises the need to modernise legacy data infrastructure as  data is a ``strategic asset'' \citep{national2020}. 

Although data infrastructure systems do not usually include complicated algorithms or artificial intelligence, the quality and consistency of the data collected are essential prerequisites for reliable analytical analysis or predictions. It is, therefore, crucial to ensure that the data collection process achieves good quality, fit for purpose data and does not unnecessarily infringe on individual privacy. To examine the application of such principles, we consider the West Midlands Police data-driven policing project with Accenture as a data infrastructure case study. 

As noted above, this data infrastructure collates various datasets on police data, including `stop and search'. Consider, for example, the way information about race and ethnicity may be collected by law enforcement in `stop and search' datasets. In the USA, data collected includes separate fields for race and ethnicity, yet not all agencies fill in the ethnicity field \citep{Fogliato_2021}. This creates a bias in the data, which can be very difficult to fix post hoc. In addition, race and ethnicity as defined by the individual might be different from the race and ethnicity observed and recorded by the arresting officer. In the UK, it is clearly stated if the race and ethnicity field is based on the individual’s statement or assumed by the officer, with preference given to the former. Further, on a societal level, where certain demographics have higher arrest rates, police may be more likely to stop and search based on observed race, compounding and reinforcing existing unfairness.\footnote{Issues of race disproportionality within policing were discussed in the Lammy Review, The Stephen Lawrence Inquiry and Race Disparity Audit, and in the discussion in \citep{CDEIreport}.} The report on the project does not discuss principles of trustworthy design \citep{accenture2020west}. The West Midlands Police and Crime Commissioner Ethics Committee raised potential bias as an issue arising from the data infrastructure set up by Accenture \citep{westmidlandsethicscommittee}. 

Data infrastructure that reinforces existing bias will not be fit for purpose or fair. The Accenture project does not clarify how the use of data from various sources is being dealt with. Specifically, different types of data have different levels of reliability. For example, intelligence reports cannot be taken on par with proven offending data. 

Law enforcement recognises that privacy and security are issues with modern data infrastructures \citep{metpolice2021}. Recent developments in privacy-enhancing technologies, including differential privacy and secure multi-party computation, can sometimes enable useful information to be discovered without inappropriate access to private data \citep{royalsociety}. 

\subsection{Data analysis}

The use of data analysis as part of decision-making processes may give rise to certain risks. As noted above, there are various data analysis tools in use across the UK law enforcement agencies. However, the Home Office recently submitted that ``our analysis suggests that most data analytics used by police forces is currently used to enable organisational effectiveness and resource planning rather than directly to tackle crime'' \citep{homeoffice2021}. It appears there is a lack of central understanding of what algorithmic tools are in use. A lack of national standards creates concerns about fairness, transparency and accountability.

MAPinfo is an example of an analytics tool that uses past incident reports to identify crime `hotspots'. Police officers are alerted to these areas for patrols and monitoring. Although not strictly a predictive tool, careless use of hotspot mapping can still result in over-policing of local communities -- such that a level of policing ``is disproportionate to the level of crime that actually exists in a given area'' \citep{liberty2021}. This follows a similar concern as raised in respect of data collection above. If data collection is not fit for purpose, this will flow through to later stages of the data pipeline. For example, an algorithm trained on biased data will learn to repeat or even reinforce the same bias unless steps are taken to avoid it. There is potential for bias to be further aggravated by algorithmic tools – for example, via feedback loops \citep{ensign2018runaway}.

\subsection{Risk Prediction}

Predicting future crime has been the focus of much research and legal development \citep{propublica, berk2013statistical,flores2016false,fazel2021predictive}. The high-risk nature of predicting future criminal offences has induced procedural safeguards that ensure law enforcement cannot depend on entirely automated risk predictions.\footnote{\textit{Data Protection Act 2018} (UK) section 50; \textit{General Data Protection Regulations} (EU) section 22. See also draft provisions in \textit{Draft Artificial Intelligence Act 2021} (EU) section 19 and \textit{Bill for an Algorithmic Accountability Act of 2022} (USA).}

Even where such tools are not solely used to make decisions, further consideration should be given to fairness generally in risk prediction and ensure it is not disproportionate to individual rights and freedoms. Risk predictions about a persons' future behaviour may have long-lasting ramifications. Unlike the other two categorisations of algorithmic tools, this type of tool may be relevant to court proceedings. Transparency is a prerequisite for individuals to understand the way algorithmic tools are designed and used, a necessity  to be able to challenge this use. However, the inference rules  algorithmic risk prediction models `learn' when training may not be easily understood by humans \citep{machinelearningvestby}.   

HM Prison and Probation Service have advanced a policy to ``deploy an advanced algorithm which draws on data from the whole of the prisoner’s journey through the criminal justice system to produce a recommended initial categorisation'' \citep{prisonreformtrust2021,prisoncategorisationpolicy2021}. This categorisation is reviewed and can be changed by the staff. The Prison Reform Trust, an independent UK charity, highlights the welcome prospects of ``more consistent application of the criteria that determine categorisation'' \citep{prisonreformtrust2021}. However, the quality of the prediction depends on the quality of information that is fed into the system which cannot easily be validated or challenged. This also puts in question the ability of the staff to effectively challenge the categorisation suggested by the tool. The responsibility of ensuring the tool works properly falls on a manager appointed by the Prison Governor. Specifically, the manager is required to ``that the categorisation/recategorisation process is functioning effectively; that decisions are fair, consistent and taken without bias; to provide quality assurance of decision making; to collect and analyse data in terms of protected characteristics alongside other equalities data to ensure that there is a complete picture of any disproportionate impact, and to implement change where necessary'' \citep{prisonreformtrust2021}. However, it is unlikely local managers are equipped to do this. 

\subsection{Implication of the deployment of multiple algorithms simultaneously}

The UK criminal justice system is not unusual in having many players and components, such as regional law enforcement agencies. Ensuring trustworthy deployment within the criminal justice system requires considering the deployment of other algorithmic tools in the same jurisdiction. Specifically, the deployment of algorithmic systems in earlier stages of the pipeline, i.e., policing, can introduce both noise and bias to algorithms deployed later within the pipeline, e.g., within the courts. Consider, for example, tools that predict risk associated with an individual, like those used in the pre-sentencing report in the UK (see Section~\ref{sec:alg_court}). These tools often take into account the number of prior arrests in determining the level of risk. Even if those algorithms are well designed and thoroughly tested, they may implicitly depend on the assumption that an individual's likelihood of arrest is independent of, for example, the neighbourhood where they reside. If this stops being the case at any point of deployment, potentially due to a non-reliable deployment of a predictive policing algorithm within one or more jurisdictions, this may affect the reliability of the individualised risk prediction tool used in the courts. 

\subsection{Ethics, governance and regulation}
Law enforcement agencies have a positive duty to ``satisfy themselves, either directly or by way of independent verification, that the software program in this case does not have an unacceptable bias on the grounds of race or sex''.\footnote{\textit{Bridges v South Wales Police} para 199.} However, it is not clear that individual agencies have the resources, skills and understanding to verify an absence of bias. For example, merely removing protected characteristics from data analytics tools is not sufficient \citep{Joshua2021, homeoffice2021}. There is often a need for law enforcement to find commercial partners to develop algorithmic tools. However, undertaking verification of an absence of bias may be particularly difficult when using proprietary software. The Metropolitan Police have highlighted ``vendors being reluctant to share enough information citing reasons of commercial confidentiality'' as a challenge \citep{metpolice2021}. It is important to remember that even where vendors claim commercial confidentiality over algorithmic tools, a public authority cannot neglect its non-delegable Public Sector Equality Duty.\footnote{\textit{Bridges v South Wales Police} para 199, see Public Sector Equality Duty in \textit{Equality Act 2010} section 149.}

The Home Office states that ``we strongly disagree with the characterisation put forward by some that these technologies are being deployed without regard to proper evaluation or engagement, or that we are allowing sensitive decisions to be delegated to machines in a way that is either contrary to the law or the core principles of the CJS (criminal justice system)'' \citep{Home_Office}. However, it is unclear what these proper evaluations are and who is in charge of ensuring they are properly executed. From the evidence provided by Home Office and other law enforcement agencies, it is difficult to establish the governance structure with respect to ensuing proper evaluation of new technologies. For example, both Kent Police and the Metropolitan Police have previously deployed and later abandoned the predictive policing software \textit{Predpol} \citep{Predpol_kent, Predpol_met}, a tool still widely used in the US, with strong claims that its use can potentially increase racial bias \citep{sankin_2021_crime}. 

Although it is often stated that all algorithmic decisions are advisory and that accountability remains with the practitioners, this is only typically valuable 
when practitioners are in a position to effectively verify or challenge the decision, which is often not the case. Furthermore, the decision to incorporate tools is often not in the hands of individual officers or practitioners, adding to the limits of holding them accountable for mistakes caused by using the tool. The Metropolitan Police highlight the challenge of internally providing effective oversight for the systems: ``Policing would welcome the support from experts in Government and beyond to help meet this challenge''. Specifically, they highlighted ``A code of practice would provide a framework for ethical decision making when considering whether to use a new technology'' \citep{metpolice2021}. We note that there are existing frameworks available \citep{Oswald2018, fazel2018}, and that a national framework for policing is being developed by the National Police Chiefs’ Council and the Centre for Data Ethics and Innovation. 

Currently, several police forces take their data analytics products to independent ethics committees and scrutiny panels prior to deployment. These are often comprised of volunteers and operate in an advisory capacity. Meeting notes are typically made public and can be further scrutinised by interested parties. However, projects are often considered by ethics committees after much time and resources have been invested in them. As such, there may be an implicit pressure to approve or to find post hoc solutions to any problems found. Governance structures are needed to ensure that teams deploying data analytics have the knowledge and resources to intentionally design trustworthy systems from the beginning. Demonstrating an intentionally trustworthy design is likely to both save resources and provide better results. In addition, inter-agency collaboration is important to ensure a consistent approach toward ethics and evaluation of advanced algorithms. A national Data Ethics Governance model, which the Home Office stated is planned, is needed \citep{Home_Office}. It is crucial to ensure that governance structures allocate responsibilities to those with the resources, support and ability to undertake appropriate due diligence \citep{prisonreformtrust2021}. 

\section{Discussion}
\label{sec:discussion}

Data-driven technologies are increasingly  an important part of the criminal justice system. Avon and Somerset Constabulary stated  they ``believe that data is a critical organisational asset and, like our staff, is at the core of delivering efficient and effective local policing services'' \citep{Avon2021}. The Metropolitan Police stated that ``To declare technologies as being ‘off limits’ to policing risks denying law enforcement the tools it needs to keep the public safe whilst leaving these tools easily available for criminals and commercial users to consume and exploit'' \citep{metpolice2021}. The Home Office endorses the use of technology to improve effectiveness and efficiency within the criminal justice system \citep{Home_Office2021}.

While specific tools such as \textit{Predpol} and \textit{COMPAS} have received significant attention from the algorithmic fairness community, there are many more algorithmic systems trialled and deployed around the world. In this work, we use the UK as a case study to step back and consider the deployment of algorithmic tools at different decision points in the criminal justice system. We consider how different trustworthy design principles might apply to examples of tools in deployment and examine the current state of guidance and regulation of these tools in deployment. Although ethical decision making and evaluation frameworks are available and more are in development, it is unclear who is able and accountable to perform the required due diligence. It is implied that: the practitioner is responsible for validating or challenging the automated decision; local management is responsible for ensuring the tool is fit for purpose, free of bias and is working as expected; and in some cases, it is the responsibility of advisory ethics committees, composed of volunteers, to ethically scrutinise and approve the use of operational systems. Unfortunately, it is often the case that none of the parties above has the ability in terms of access to data, skill, time and resources to perform the required level of due diligence. 

We conclude with the following proposals: (1) A clear governance structure is required, identifying an appropriate body able to assist agencies in choosing or building trustworthy tools and systems, and ensuring proper testing and due diligence occur prior to deployment. (2) Outcomes of testing and evaluation of new tools should be made public unless security prohibits this, and mechanisms for independent scrutiny of the systems should be made available wherever possible (for example, by academic-led teams). 

The House of Lords Justice and Home Affairs Committee published its report on \textit{Technology Rules? The advent of new technologies in the justice system} \citep{MOJ_2022}.
The Committee examined the use of algorithmic tools throughout the \textit{criminal justice pipeline} and proposed many recommendations that touch on issues identified in this paper, including, but not limited to: legislation establishing clear principles and standards for the use of technologies in the justice system (recommendations 9 and 12); establishing a proper governance structure for the use of technologies in the application of the law (recommendation 5); the need for an independent body to support the various entities involved in designing and deploying new technologies (recommendation 6);  empowering local specialist ethics committees in the law enforcement community (recommendation 35). In addition, the Committee concludes that humans should always be the ultimate decision-maker and that there should be more clarity on the possible duty of candour for police to ensure appropriate transparency over their use of AI (recommendation 18). 

We welcome these recommendations and hope that further research, law reform and governance oversight will bring these recommendations into effective practice.
%

\begin{acks}
The authors thank Detective Sergeant Laurence Cartwright for valuable comments and discussion. We thank the reviewers for their helpful comments and suggestions. M.Z. acknowledges support from EPSRC grant EP/V025279/1, The Alan Turing Institute and the Leverhulme Trust grant ECF-2021-429. H.S. acknowledges support from the General Sir John Monash Foundation. A.W. acknowledges support from a Turing AI Fellowship under grant EP/V025279/1, The Alan Turing Institute, and the Leverhulme Trust via CFI.
\end{acks}


\bibliographystyle{ACM-Reference-Format}
\bibliography{main}


\begin{thebibliography}{82}


\ifx \showCODEN    \undefined \def \showCODEN     #1{\unskip}     \fi
\ifx \showDOI      \undefined \def \showDOI       #1{#1}\fi
\ifx \showISBNx    \undefined \def \showISBNx     #1{\unskip}     \fi
\ifx \showISBNxiii \undefined \def \showISBNxiii  #1{\unskip}     \fi
\ifx \showISSN     \undefined \def \showISSN      #1{\unskip}     \fi
\ifx \showLCCN     \undefined \def \showLCCN      #1{\unskip}     \fi
\ifx \shownote     \undefined \def \shownote      #1{#1}          \fi
\ifx \showarticletitle \undefined \def \showarticletitle #1{#1}   \fi
\ifx \showURL      \undefined \def \showURL       {\relax}        \fi
\providecommand\bibfield[2]{#2}
\providecommand\bibinfo[2]{#2}
\providecommand\natexlab[1]{#1}
\providecommand\showeprint[2][]{arXiv:#2}

\bibitem[\protect\citeauthoryear{Accenture}{Accenture}{2020}]%
        {accenture2020west}
\bibfield{author}{\bibinfo{person}{Accenture}.}
  \bibinfo{year}{2020}\natexlab{}.
\newblock \bibinfo{title}{West Midlands Police: Serving and protecting with
  cloud}.
\newblock
\newblock
\urldef\tempurl%
\url{https://www.accenture.com/_acnmedia/PDF-142/Accenture-West-Midlands-Police-Serving-and-protecting-with-cloud.pdf#zoom=40}
\showURL{%
\tempurl}


\bibitem[\protect\citeauthoryear{{Australian Human Rights
  Commission}}{{Australian Human Rights Commission}}{2021}]%
        {australian_human_rights_commission_human_2021}
\bibfield{author}{\bibinfo{person}{{Australian Human Rights Commission}}.}
  \bibinfo{year}{2021}\natexlab{}.
\newblock \bibinfo{booktitle}{\emph{Human Rights and Technology}}.
\newblock \bibinfo{type}{Final Report}. \bibinfo{institution}{Australian Human
  Rights Commission}.
\newblock


\bibitem[\protect\citeauthoryear{Avon and Police}{Avon and Police}{2021}]%
        {Avon2021}
\bibfield{author}{\bibinfo{person}{Avon} {and} \bibinfo{person}{Somerset
  Police}.} \bibinfo{year}{2021}\natexlab{}.
\newblock \bibinfo{title}{Avon and Somerset Police — Written evidence
  (NTL0052)}.
\newblock
\newblock
\urldef\tempurl%
\url{https://committees.parliament.uk/writtenevidence/40328/pdf/}
\showURL{%
\tempurl}


\bibitem[\protect\citeauthoryear{Babuta and Oswald}{Babuta and Oswald}{2018}]%
        {RUSI}
\bibfield{author}{\bibinfo{person}{Alexander Babuta} {and}
  \bibinfo{person}{Marion Oswald}.} \bibinfo{year}{2018}\natexlab{}.
\newblock \bibinfo{booktitle}{\emph{Machine Learning Algorithms and Police
  Decision-Making: Legal, Ethical and Regulatory Challenges}}.
\newblock \bibinfo{type}{{T}echnical {R}eport}. \bibinfo{institution}{Royal
  United Services Institute}.
\newblock
\urldef\tempurl%
\url{https://rusi.org/explore-our-research/publications/whitehall-reports/machine-learning-algorithms-and-police-decision-making-legal-ethical-and-regulatory-challenges}
\showURL{%
\tempurl}


\bibitem[\protect\citeauthoryear{Babuta and Oswald}{Babuta and Oswald}{2020}]%
        {babuta2020data}
\bibfield{author}{\bibinfo{person}{Alexander Babuta} {and}
  \bibinfo{person}{Marion Oswald}.} \bibinfo{year}{2020}\natexlab{}.
\newblock \bibinfo{title}{Data analytics and algorithms in policing in England
  and Wales: Towards a new policy framework}.
\newblock
\newblock


\bibitem[\protect\citeauthoryear{Berk and Bleich}{Berk and Bleich}{2013}]%
        {berk2013statistical}
\bibfield{author}{\bibinfo{person}{Richard~A Berk} {and}
  \bibinfo{person}{Justin Bleich}.} \bibinfo{year}{2013}\natexlab{}.
\newblock \showarticletitle{Statistical procedures for forecasting criminal
  behavior: A comparative assessment}.
\newblock \bibinfo{journal}{\emph{Criminology \& Pub. Pol'y}}
  \bibinfo{volume}{12} (\bibinfo{year}{2013}), \bibinfo{pages}{513}.
\newblock


\bibitem[\protect\citeauthoryear{Bhatt, Andrus, Weller, and Xiang}{Bhatt
  et~al\mbox{.}}{2020}]%
        {bhatt2020machine}
\bibfield{author}{\bibinfo{person}{Umang Bhatt}, \bibinfo{person}{McKane
  Andrus}, \bibinfo{person}{Adrian Weller}, {and} \bibinfo{person}{Alice
  Xiang}.} \bibinfo{year}{2020}\natexlab{}.
\newblock \showarticletitle{Machine learning explainability for external
  stakeholders}.
\newblock \bibinfo{journal}{\emph{arXiv preprint arXiv:2007.05408}}
  (\bibinfo{year}{2020}).
\newblock


\bibitem[\protect\citeauthoryear{Cheng}{Cheng}{2021}]%
        {Cheng_2012}
\bibfield{author}{\bibinfo{person}{Tao Cheng}.}
  \bibinfo{year}{2021}\natexlab{}.
\newblock \bibinfo{title}{Crime, Policing and Citizenship (CPC) - Space-Time
  Interactions of Dynamic Networks}.
\newblock
\newblock
\urldef\tempurl%
\url{https://gtr.ukri.org/projects?ref=EP%2FJ004197%2F1}
\showURL{%
\tempurl}
\newblock
\shownote{Accessed: 2021}.


\bibitem[\protect\citeauthoryear{Chouldechova}{Chouldechova}{2017}]%
        {chouldechova2017fair}
\bibfield{author}{\bibinfo{person}{Alexandra Chouldechova}.}
  \bibinfo{year}{2017}\natexlab{}.
\newblock \showarticletitle{Fair prediction with disparate impact: A study of
  bias in recidivism prediction instruments}.
\newblock \bibinfo{journal}{\emph{Big data}} \bibinfo{volume}{5},
  \bibinfo{number}{2} (\bibinfo{year}{2017}), \bibinfo{pages}{153--163}.
\newblock


\bibitem[\protect\citeauthoryear{Coyle and Weller}{Coyle and Weller}{2020}]%
        {coyle2020explaining}
\bibfield{author}{\bibinfo{person}{Diane Coyle} {and} \bibinfo{person}{Adrian
  Weller}.} \bibinfo{year}{2020}\natexlab{}.
\newblock \showarticletitle{“Explaining” machine learning reveals policy
  challenges}.
\newblock \bibinfo{journal}{\emph{Science}} \bibinfo{volume}{368},
  \bibinfo{number}{6498} (\bibinfo{year}{2020}), \bibinfo{pages}{1433--1434}.
\newblock


\bibitem[\protect\citeauthoryear{Dencik, Hintz, Redden, and Warne}{Dencik
  et~al\mbox{.}}{2018}]%
        {dencik2018}
\bibfield{author}{\bibinfo{person}{Lina Dencik}, \bibinfo{person}{Arne Hintz},
  \bibinfo{person}{Joanna Redden}, {and} \bibinfo{person}{Harry Warne}.}
  \bibinfo{year}{2018}\natexlab{}.
\newblock \bibinfo{title}{Data scores as governance: Investigating uses of
  citizen scoring in public services project report}.
\newblock
\newblock


\bibitem[\protect\citeauthoryear{Devon and Police}{Devon and Police}{2020}]%
        {STORM_Devon_and_Cornwall_Police}
\bibfield{author}{\bibinfo{person}{Devon} {and} \bibinfo{person}{Cornwall
  Police}.} \bibinfo{year}{2020}\natexlab{}.
\newblock \bibinfo{title}{Force Call Handling and Contact Policy}.
\newblock
\newblock
\urldef\tempurl%
\url{https://www.devon-cornwall.police.uk/FOI/Doc/d4999792-4609-4f1c-aba2-2948f7bba40a/p?D268open.pdf}
\showURL{%
\tempurl}


\bibitem[\protect\citeauthoryear{Ensign, Friedler, Neville, Scheidegger, and
  Venkatasubramanian}{Ensign et~al\mbox{.}}{2018}]%
        {ensign2018runaway}
\bibfield{author}{\bibinfo{person}{Danielle Ensign}, \bibinfo{person}{Sorelle~A
  Friedler}, \bibinfo{person}{Scott Neville}, \bibinfo{person}{Carlos
  Scheidegger}, {and} \bibinfo{person}{Suresh Venkatasubramanian}.}
  \bibinfo{year}{2018}\natexlab{}.
\newblock \showarticletitle{Runaway feedback loops in predictive policing}. In
  \bibinfo{booktitle}{\emph{Conference on Fairness, Accountability and
  Transparency}}. \bibinfo{publisher}{PMLR}, \bibinfo{pages}{160--171}.
\newblock


\bibitem[\protect\citeauthoryear{Eubanks}{Eubanks}{2018}]%
        {eubanks_automating_2018}
\bibfield{author}{\bibinfo{person}{Virginia Eubanks}.}
  \bibinfo{year}{2018}\natexlab{}.
\newblock \bibinfo{booktitle}{\emph{Automating Inequality}}.
\newblock \bibinfo{publisher}{St. Martin's Press}.
\newblock


\bibitem[\protect\citeauthoryear{Fazel, Burghart, Fanshawe, Gil, Monahan, and
  Yu}{Fazel et~al\mbox{.}}{2021}]%
        {fazel2021predictive}
\bibfield{author}{\bibinfo{person}{Seena Fazel}, \bibinfo{person}{Matthias
  Burghart}, \bibinfo{person}{Thomas Fanshawe},
  \bibinfo{person}{Sharon~Danielle Gil}, \bibinfo{person}{John Monahan}, {and}
  \bibinfo{person}{Rongqin Yu}.} \bibinfo{year}{2021}\natexlab{}.
\newblock \bibinfo{title}{The predictive performance of criminal risk
  assessment tools used at sentencing}.
\newblock
\newblock


\bibitem[\protect\citeauthoryear{Fazel and Wolf}{Fazel and Wolf}{2018}]%
        {fazel2018}
\bibfield{author}{\bibinfo{person}{Seena Fazel} {and} \bibinfo{person}{Achim
  Wolf}.} \bibinfo{year}{2018}\natexlab{}.
\newblock \showarticletitle{Selecting a risk assessment tool to use in
  practice: a 10-point guide}.
\newblock \bibinfo{journal}{\emph{Evidence-based mental health}}
  \bibinfo{volume}{21}, \bibinfo{number}{2} (\bibinfo{year}{2018}),
  \bibinfo{pages}{41--43}.
\newblock


\bibitem[\protect\citeauthoryear{Flores, Bechtel, and Lowenkamp}{Flores
  et~al\mbox{.}}{2016}]%
        {flores2016false}
\bibfield{author}{\bibinfo{person}{Anthony~W Flores}, \bibinfo{person}{Kristin
  Bechtel}, {and} \bibinfo{person}{Christopher~T Lowenkamp}.}
  \bibinfo{year}{2016}\natexlab{}.
\newblock \showarticletitle{False positives, false negatives, and false
  analyses: A rejoinder to machine bias: There's software used across the
  country to predict future criminals. and it's biased against blacks}.
\newblock \bibinfo{journal}{\emph{Fed. Probation}}  \bibinfo{volume}{80}
  (\bibinfo{year}{2016}), \bibinfo{pages}{38}.
\newblock


\bibitem[\protect\citeauthoryear{Fogliato, Xiang, Lipton, Nagin, and
  Chouldechova}{Fogliato et~al\mbox{.}}{2021}]%
        {Fogliato_2021}
\bibfield{author}{\bibinfo{person}{Riccardo Fogliato}, \bibinfo{person}{Alice
  Xiang}, \bibinfo{person}{Zachary Lipton}, \bibinfo{person}{Daniel Nagin},
  {and} \bibinfo{person}{Alexandra Chouldechova}.}
  \bibinfo{year}{2021}\natexlab{}.
\newblock \showarticletitle{On the Validity of Arrest as a Proxy for Offense:
  Race and the Likelihood of Arrest for Violent Crimes}. In
  \bibinfo{booktitle}{\emph{Proceedings of the 2021 {AAAI}/{ACM} Conference on
  {AI}, Ethics, and Society}}. \bibinfo{publisher}{{ACM}}.
\newblock
\urldef\tempurl%
\url{https://doi.org/10.1145/3461702.3462538}
\showDOI{\tempurl}


\bibitem[\protect\citeauthoryear{for Data~Ethics and Innovation}{for
  Data~Ethics and Innovation}{2021}]%
        {CDEIreport}
\bibfield{author}{\bibinfo{person}{Centre for Data~Ethics} {and}
  \bibinfo{person}{Innovation}.} \bibinfo{year}{2021}\natexlab{}.
\newblock \bibinfo{title}{Review into bias in algorithmic decision-making}.
\newblock
\newblock


\bibitem[\protect\citeauthoryear{Howard and Dixon}{Howard and Dixon}{2013}]%
        {oasys}
\bibfield{author}{\bibinfo{person}{Philip~D Howard} {and}
  \bibinfo{person}{Louise Dixon}.} \bibinfo{year}{2013}\natexlab{}.
\newblock \showarticletitle{Identifying change in the likelihood of violent
  recidivism: Causal dynamic risk factors in the OASys violence predictor.}
\newblock \bibinfo{journal}{\emph{Law and Human Behavior}}
  \bibinfo{volume}{37}, \bibinfo{number}{3} (\bibinfo{year}{2013}),
  \bibinfo{pages}{163}.
\newblock


\bibitem[\protect\citeauthoryear{Inspectorates}{Inspectorates}{2019a}]%
        {THRIVE}
\bibfield{author}{\bibinfo{person}{Criminal~Justice Inspectorates}.}
  \bibinfo{year}{2019}\natexlab{a}.
\newblock \bibinfo{title}{DASH}.
\newblock
\newblock
\urldef\tempurl%
\url{https://www.justiceinspectorates.gov.uk/hmicfrs/glossary/thrive/}
\showURL{%
\tempurl}


\bibitem[\protect\citeauthoryear{Inspectorates}{Inspectorates}{2019b}]%
        {DASH}
\bibfield{author}{\bibinfo{person}{Criminal~Justice Inspectorates}.}
  \bibinfo{year}{2019}\natexlab{b}.
\newblock \bibinfo{title}{DASH}.
\newblock
\newblock
\urldef\tempurl%
\url{https://www.justiceinspectorates.gov.uk/hmicfrs/glossary/dash}
\showURL{%
\tempurl}


\bibitem[\protect\citeauthoryear{International}{International}{2018}]%
        {amnesty2018}
\bibfield{author}{\bibinfo{person}{Amnesty International}.}
  \bibinfo{year}{2018}\natexlab{}.
\newblock \bibinfo{title}{Trapped in the Matrix}.
\newblock
\newblock
\urldef\tempurl%
\url{https://www.amnesty.org.uk/files/reports/Trapped%20in%20the%20Matrix%20Amnesty%20report.pdf}
\showURL{%
\tempurl}


\bibitem[\protect\citeauthoryear{Jansen}{Jansen}{2018}]%
        {jansen2018data}
\bibfield{author}{\bibinfo{person}{Fieke Jansen}.}
  \bibinfo{year}{2018}\natexlab{}.
\newblock \showarticletitle{Data Driven Policing in the Context of Europe}.
\newblock \bibinfo{journal}{\emph{Data Justice Lab}} (\bibinfo{year}{2018}).
\newblock


\bibitem[\protect\citeauthoryear{Justice and Committee}{Justice and
  Committee}{2022}]%
        {MOJ_2022}
\bibfield{author}{\bibinfo{person}{Justice} {and} \bibinfo{person}{Home~Affairs
  Committee}.} \bibinfo{year}{2022}\natexlab{}.
\newblock \bibinfo{title}{Technology Rules? The advent of new technologies in
  the justice system}.
\newblock
\newblock
\urldef\tempurl%
\url{https://publications.parliament.uk/pa/ld5802/ldselect/ldjusthom/180/18002.htm}
\showURL{%
\tempurl}


\bibitem[\protect\citeauthoryear{Liberty}{Liberty}{2021}]%
        {liberty2021}
\bibfield{author}{\bibinfo{person}{Liberty}.} \bibinfo{year}{2021}\natexlab{}.
\newblock \bibinfo{title}{New technologies and the application of the law:
  Written evidence}.
\newblock
\newblock
\urldef\tempurl%
\url{https://committees.parliament.uk/writtenevidence/38701/pdf/}
\showURL{%
\tempurl}


\bibitem[\protect\citeauthoryear{Lum and Isaac}{Lum and Isaac}{2016}]%
        {lum2016predict}
\bibfield{author}{\bibinfo{person}{Kristian Lum} {and} \bibinfo{person}{William
  Isaac}.} \bibinfo{year}{2016}\natexlab{}.
\newblock \showarticletitle{To predict and serve?}
\newblock \bibinfo{journal}{\emph{Significance}} \bibinfo{volume}{13},
  \bibinfo{number}{5} (\bibinfo{year}{2016}), \bibinfo{pages}{14--19}.
\newblock


\bibitem[\protect\citeauthoryear{Lyons, Velloso, and Miller}{Lyons
  et~al\mbox{.}}{2021}]%
        {lyons2021fair}
\bibfield{author}{\bibinfo{person}{Henrietta Lyons}, \bibinfo{person}{Eduardo
  Velloso}, {and} \bibinfo{person}{Tim Miller}.}
  \bibinfo{year}{2021}\natexlab{}.
\newblock \showarticletitle{Fair and Responsible AI: A focus on the ability to
  contest}.
\newblock \bibinfo{journal}{\emph{arXiv preprint arXiv:2102.10787}}
  (\bibinfo{year}{2021}).
\newblock


\bibitem[\protect\citeauthoryear{Mamalian and La~Vigne}{Mamalian and
  La~Vigne}{1999}]%
        {mamalian1999use}
\bibfield{author}{\bibinfo{person}{Cynthia~A Mamalian} {and}
  \bibinfo{person}{Nancy~Gladys La~Vigne}.} \bibinfo{year}{1999}\natexlab{}.
\newblock \bibinfo{title}{The use of computerized crime mapping by law
  enforcement: Survey results}.
\newblock
\newblock


\bibitem[\protect\citeauthoryear{McFadzien, Pughsley, Featherstone, and
  Phillips}{McFadzien et~al\mbox{.}}{2020}]%
        {mcfadzien2020evidence}
\bibfield{author}{\bibinfo{person}{Kent McFadzien}, \bibinfo{person}{Alan
  Pughsley}, \bibinfo{person}{Andrew~M Featherstone}, {and}
  \bibinfo{person}{John~M Phillips}.} \bibinfo{year}{2020}\natexlab{}.
\newblock \showarticletitle{The evidence-based investigative tool (EBIT): a
  legitimacy-conscious statistical triage process for high-volume crimes}.
\newblock \bibinfo{journal}{\emph{Cambridge Journal of Evidence-Based
  Policing}} \bibinfo{volume}{4}, \bibinfo{number}{3} (\bibinfo{year}{2020}),
  \bibinfo{pages}{218--232}.
\newblock


\bibitem[\protect\citeauthoryear{Mulligan, Kluttz, and Kohli}{Mulligan
  et~al\mbox{.}}{2019}]%
        {mulligan2019shaping}
\bibfield{author}{\bibinfo{person}{Deirdre~K Mulligan}, \bibinfo{person}{Daniel
  Kluttz}, {and} \bibinfo{person}{Nitin Kohli}.}
  \bibinfo{year}{2019}\natexlab{}.
\newblock \showarticletitle{Shaping our tools: Contestability as a means to
  promote responsible algorithmic decision making in the professions}.
\newblock \bibinfo{journal}{\emph{Available at SSRN 3311894}}
  (\bibinfo{year}{2019}).
\newblock


\bibitem[\protect\citeauthoryear{Newsroom}{Newsroom}{2017}]%
        {Newsroom}
\bibfield{author}{\bibinfo{person}{The Newsroom}.}
  \bibinfo{year}{2017}\natexlab{}.
\newblock \bibinfo{title}{I predict a break-in: Yorkshire police use
  cutting-edge technology to deter burglars}.
\newblock
\newblock
\urldef\tempurl%
\url{https://www.yorkshirepost.co.uk/news/crime/i-predict-break-yorkshire-police-use-cutting-edge-technology-deter-burglars-595904}
\showURL{%
\tempurl}


\bibitem[\protect\citeauthoryear{Nilsson}{Nilsson}{2018}]%
        {Predpol_met}
\bibfield{author}{\bibinfo{person}{Patricia Nilsson}.}
  \bibinfo{year}{2018}\natexlab{}.
\newblock \bibinfo{title}{First UK police force to try predictive policing ends
  contract}.
\newblock
\newblock
\urldef\tempurl%
\url{https://www.ft.com/content/b34b0b08-ef19-11e8-89c8-d36339d835c0}
\showURL{%
\tempurl}
\newblock
\shownote{Accessed: 2022}.


\bibitem[\protect\citeauthoryear{(NPCC), of~Police, and Commissioners}{(NPCC)
  et~al\mbox{.}}{2020}]%
        {national2020}
\bibfield{author}{\bibinfo{person}{The National Police Chiefs~Council (NPCC)},
  \bibinfo{person}{Association of Police}, {and} \bibinfo{person}{Crime
  Commissioners}.} \bibinfo{year}{2020}\natexlab{}.
\newblock \bibinfo{title}{National Policing Digital Strategy 2020-2030}.
\newblock
\newblock
\urldef\tempurl%
\url{https://www.apccs.police.uk/media/4886/national-policing-digital-strategy-2020-2030.pdf}
\showURL{%
\tempurl}


\bibitem[\protect\citeauthoryear{of~Justice}{of~Justice}{2021}]%
        {MOJ}
\bibfield{author}{\bibinfo{person}{The~Ministry of Justice}.}
  \bibinfo{year}{2021}\natexlab{}.
\newblock \bibinfo{title}{Ministry of Justice — Written evidence (NTL0053)}.
\newblock
\newblock
\urldef\tempurl%
\url{https://committees.parliament.uk/writtenevidence/40365/pdf/}
\showURL{%
\tempurl}


\bibitem[\protect\citeauthoryear{Office}{Office}{[n.\,d.]}]%
        {PNC}
\bibfield{author}{\bibinfo{person}{ACRO Criminal~Records Office}.}
  \bibinfo{year}{[n.\,d.]}\natexlab{}.
\newblock \bibinfo{title}{Police National Computer Services}.
\newblock
\newblock
\urldef\tempurl%
\url{https://www.acro.police.uk/PNC-services}
\showURL{%
\tempurl}
\newblock
\shownote{Accessed: 2022}.


\bibitem[\protect\citeauthoryear{Office}{Office}{2021a}]%
        {homeoffice2021}
\bibfield{author}{\bibinfo{person}{Home Office}.}
  \bibinfo{year}{2021}\natexlab{a}.
\newblock \bibinfo{title}{New technologies and the application of the law:
  Written evidence}.
\newblock
\newblock
\urldef\tempurl%
\url{https://committees.parliament.uk/writtenevidence/40975/pdf/}
\showURL{%
\tempurl}


\bibitem[\protect\citeauthoryear{Office}{Office}{2021b}]%
        {Home_Office}
\bibfield{author}{\bibinfo{person}{The~Home Office}.}
  \bibinfo{year}{2021}\natexlab{b}.
\newblock \bibinfo{title}{Home Office — Written evidence (NTL0055)}.
\newblock
\newblock
\urldef\tempurl%
\url{https://committees.parliament.uk/writtenevidence/40975/pdf/}
\showURL{%
\tempurl}


\bibitem[\protect\citeauthoryear{Oswald, Grace, Urwin, and Barnes}{Oswald
  et~al\mbox{.}}{2018}]%
        {Oswald2018}
\bibfield{author}{\bibinfo{person}{Marion Oswald}, \bibinfo{person}{Jamie
  Grace}, \bibinfo{person}{Sheena Urwin}, {and} \bibinfo{person}{Geoffrey~C.
  Barnes}.} \bibinfo{year}{2018}\natexlab{}.
\newblock \showarticletitle{Algorithmic risk assessment policing models:
  lessons from the Durham HART model and ‘Experimental’ proportionality}.
\newblock \bibinfo{journal}{\emph{Information \& Communications Technology
  Law}} \bibinfo{volume}{27}, \bibinfo{number}{2} (\bibinfo{year}{2018}),
  \bibinfo{pages}{223--250}.
\newblock
\urldef\tempurl%
\url{https://doi.org/10.1080/13600834.2018.1458455}
\showDOI{\tempurl}


\bibitem[\protect\citeauthoryear{Pasquale}{Pasquale}{2019}]%
        {pasquale_black_2019}
\bibfield{author}{\bibinfo{person}{Frank Pasquale}.}
  \bibinfo{year}{2019}\natexlab{}.
\newblock \bibinfo{booktitle}{\emph{The Black Box Society}}.
\newblock \bibinfo{publisher}{Harvard University Press},
  \bibinfo{address}{Cambridge, MA}.
\newblock


\bibitem[\protect\citeauthoryear{Ploug and Holm}{Ploug and Holm}{2020}]%
        {ploug2020four}
\bibfield{author}{\bibinfo{person}{Thomas Ploug} {and}
  \bibinfo{person}{S{\o}ren Holm}.} \bibinfo{year}{2020}\natexlab{}.
\newblock \showarticletitle{The four dimensions of contestable AI diagnostics-A
  patient-centric approach to explainable AI}.
\newblock \bibinfo{journal}{\emph{Artificial Intelligence in Medicine}}
  \bibinfo{volume}{107} (\bibinfo{year}{2020}), \bibinfo{pages}{101901}.
\newblock


\bibitem[\protect\citeauthoryear{Police}{Police}{2020a}]%
        {STORM_Dorset_Police}
\bibfield{author}{\bibinfo{person}{Dorset Police}.}
  \bibinfo{year}{2020}\natexlab{a}.
\newblock \bibinfo{title}{Freedom of Information Act Request No: 2020-799}.
\newblock
\newblock
\urldef\tempurl%
\url{https://www.dorset.police.uk/media/64900/record-1-2020-396.doc}
\showURL{%
\tempurl}


\bibitem[\protect\citeauthoryear{Police}{Police}{2020b}]%
        {STORM_Essex_Police}
\bibfield{author}{\bibinfo{person}{Essex Police}.}
  \bibinfo{year}{2020}\natexlab{b}.
\newblock \bibinfo{title}{D0503 Procedure - Responding to Incidents}.
\newblock
\newblock
\urldef\tempurl%
\url{https://www.essex.police.uk/foi-ai/essex-police/our-policies-and-procedures/d/d0503-procedure---responding-to-incidents/}
\showURL{%
\tempurl}


\bibitem[\protect\citeauthoryear{Police}{Police}{2020c}]%
        {STORM_Leicestershire_Police}
\bibfield{author}{\bibinfo{person}{Leicestershire Police}.}
  \bibinfo{year}{2020}\natexlab{c}.
\newblock \bibinfo{title}{Freedom of Information 003528/20}.
\newblock
\newblock
\urldef\tempurl%
\url{https://www.leics.police.uk/SysSiteAssets/foi-media/leicestershire/disclosure_2020/11.-november/3528-20-iccs-and-cad-systems.pdf}
\showURL{%
\tempurl}


\bibitem[\protect\citeauthoryear{Police}{Police}{2019a}]%
        {STORM_Sussex_Police}
\bibfield{author}{\bibinfo{person}{Sussex Police}.}
  \bibinfo{year}{2019}\natexlab{a}.
\newblock \bibinfo{title}{Crime and incident disposal recording and auditing
  policy}.
\newblock
\newblock
\urldef\tempurl%
\url{https://www.sussex.police.uk/SysSiteAssets/foi-media/sussex/policies/crime-and-incident-disposal-recording-and-auditing-policy-7572019.pdf}
\showURL{%
\tempurl}


\bibitem[\protect\citeauthoryear{Police}{Police}{2020d}]%
        {STORM_Staffordshire_Police}
\bibfield{author}{\bibinfo{person}{Staffordshire Police}.}
  \bibinfo{year}{2020}\natexlab{d}.
\newblock \bibinfo{title}{Freedom of Information request: reference 12562}.
\newblock
\newblock
\urldef\tempurl%
\url{https://www.staffordshire.police.uk/SysSiteAssets/foi-media/staffordshire/published-foi-request-responses-december-2020/foi-12562-iccs--cad-systems.pdf}
\showURL{%
\tempurl}


\bibitem[\protect\citeauthoryear{Police}{Police}{2020e}]%
        {STORM_Suffolk_Police}
\bibfield{author}{\bibinfo{person}{Suffolk Police}.}
  \bibinfo{year}{2020}\natexlab{e}.
\newblock \bibinfo{title}{Freedom of Information Request Reference No : FOI
  003542/20}.
\newblock
\newblock
\urldef\tempurl%
\url{https://www.suffolk.police.uk/sites/suffolk/files/003542-20_-_iccs_and_cad_contract.pdf}
\showURL{%
\tempurl}


\bibitem[\protect\citeauthoryear{Police}{Police}{2021}]%
        {Sussex_Police}
\bibfield{author}{\bibinfo{person}{Sussex Police}.}
  \bibinfo{year}{2021}\natexlab{}.
\newblock \bibinfo{title}{Freedom of Information ref 0162/21}.
\newblock
\newblock
\urldef\tempurl%
\url{https://www.sussex.police.uk/SysSiteAssets/foi-media/sussex/other_information/freedom-of-information---foi-0162.21-covid.pdf}
\showURL{%
\tempurl}


\bibitem[\protect\citeauthoryear{Police}{Police}{[n.\,d.]a}]%
        {matrix}
\bibfield{author}{\bibinfo{person}{The~Metropolitan Police}.}
  \bibinfo{year}{[n.\,d.]}\natexlab{a}.
\newblock \bibinfo{title}{Gangs violence matrix}.
\newblock
\newblock
\urldef\tempurl%
\url{https://www.met.police.uk/police-forces/metropolitan-police/areas/about-us/about-the-met/gangs-violence-matrix/}
\showURL{%
\tempurl}
\newblock
\shownote{Accessed: 2021}.


\bibitem[\protect\citeauthoryear{Police and Commissioner}{Police and
  Commissioner}{2021}]%
        {westmidlandsethicscommittee}
\bibfield{author}{\bibinfo{person}{West Midlands~Police Police} {and}
  \bibinfo{person}{Crime Commissioner}.} \bibinfo{year}{2021}\natexlab{}.
\newblock \bibinfo{title}{Ethics Committee Minutes and Advice}.
\newblock
\newblock
\urldef\tempurl%
\url{https://www.westmidlands-pcc.gov.uk/wp-content/uploads/2021/04/05032021-EC-Minutes-and-Advice.pdf?x86241}
\showURL{%
\tempurl}


\bibitem[\protect\citeauthoryear{Police}{Police}{[n.\,d.]b}]%
        {NDAS_Accenture}
\bibfield{author}{\bibinfo{person}{West~Yorkshire Police}.}
  \bibinfo{year}{[n.\,d.]}\natexlab{b}.
\newblock \bibinfo{title}{National Data Analytics Solution (NDAS) Privacy
  Notice}.
\newblock
\newblock
\urldef\tempurl%
\url{https://www.westyorkshire.police.uk/advice/modern-slavery/national-data-analytics-solution-ndas-privacy-notice}
\showURL{%
\tempurl}
\newblock
\shownote{Accessed: 2022}.


\bibitem[\protect\citeauthoryear{Police}{Police}{2019b}]%
        {STORM_West_Yorkshire_Police}
\bibfield{author}{\bibinfo{person}{West~Yorkshire Police}.}
  \bibinfo{year}{2019}\natexlab{b}.
\newblock \bibinfo{title}{Information and Data Management}.
\newblock
\newblock
\urldef\tempurl%
\url{https://www.westyorkshire.police.uk/sites/default/files/2019-09/information_and_data_management_q.pdf}
\showURL{%
\tempurl}


\bibitem[\protect\citeauthoryear{Police}{Police}{2020f}]%
        {IOM}
\bibfield{author}{\bibinfo{person}{West~Yorkshire Police}.}
  \bibinfo{year}{2020}\natexlab{f}.
\newblock \bibinfo{title}{Integrated Offender Management (IOM)}.
\newblock
\newblock
\urldef\tempurl%
\url{https://www.westyorkshire.police.uk/sites/default/files/2020-07/integrated_offender_management_iom.pdf}
\showURL{%
\tempurl}


\bibitem[\protect\citeauthoryear{Predpol}{Predpol}{[n.\,d.]}]%
        {Predpol_kent}
\bibfield{author}{\bibinfo{person}{Predpol}.}
  \bibinfo{year}{[n.\,d.]}\natexlab{}.
\newblock \bibinfo{title}{Kent Police Use PredPol To Prevent Violent Crime}.
\newblock
\newblock
\urldef\tempurl%
\url{https://www.predpol.com/kent-police-use-predpol-to-prevent-violent-crime/}
\showURL{%
\tempurl}
\newblock
\shownote{Accessed: 2022}.


\bibitem[\protect\citeauthoryear{ProPublica}{ProPublica}{2016}]%
        {propublica}
\bibfield{author}{\bibinfo{person}{ProPublica}.}
  \bibinfo{year}{2016}\natexlab{}.
\newblock \bibinfo{title}{Machine Bias}.
\newblock
\newblock


\bibitem[\protect\citeauthoryear{Sankin, Mehrota, Mattu, and Gilbertson}{Sankin
  et~al\mbox{.}}{2021}]%
        {sankin_2021_crime}
\bibfield{author}{\bibinfo{person}{Aaron Sankin}, \bibinfo{person}{Dhruv
  Mehrota}, \bibinfo{person}{Surya Mattu}, {and} \bibinfo{person}{Annie
  Gilbertson}.} \bibinfo{year}{2021}\natexlab{}.
\newblock \bibinfo{title}{Crime Prediction Software Promised to Be Free of
  Biases. New Data Shows It Perpetuates Them}.
\newblock
\newblock
\urldef\tempurl%
\url{https://themarkup.org/prediction-bias/2021/12/02/crime-prediction-software-promised-to-be-free-of-biases-new-data-shows-it-perpetuates-them}
\showURL{%
\tempurl}


\bibitem[\protect\citeauthoryear{Santow}{Santow}{2021}]%
        {edward_santow_intelligence_2021}
\bibfield{author}{\bibinfo{person}{Edward Santow}.}
  \bibinfo{year}{2021}\natexlab{}.
\newblock \showarticletitle{Intelligence: Coding calues into the future}.
\newblock In \bibinfo{booktitle}{\emph{The Public Square Project}},
  \bibfield{editor}{\bibinfo{person}{{Peter Lewis}} {and}
  \bibinfo{person}{Jordan Guiao}} (Eds.). \bibinfo{publisher}{Melbourne
  University Press}, \bibinfo{address}{Melbourne}.
\newblock


\bibitem[\protect\citeauthoryear{Selbst}{Selbst}{2017}]%
        {selbst2017}
\bibfield{author}{\bibinfo{person}{Andrew Selbst}.}
  \bibinfo{year}{2017}\natexlab{}.
\newblock \showarticletitle{Disparate Impact in Big Data Policing}.
\newblock \bibinfo{journal}{\emph{Georgia Law Review}}  \bibinfo{volume}{52}
  (\bibinfo{year}{2017}), \bibinfo{pages}{3373}.
\newblock
Issue 1.


\bibitem[\protect\citeauthoryear{Selbst}{Selbst}{2018}]%
        {selbst_disparate_2018}
\bibfield{author}{\bibinfo{person}{Andrew Selbst}.}
  \bibinfo{year}{2018}\natexlab{}.
\newblock \showarticletitle{Disparate Impact in Big Data Policing}.
\newblock \bibinfo{journal}{\emph{Georgia Law Review}} \bibinfo{volume}{52},
  \bibinfo{number}{1} (\bibinfo{year}{2018}), \bibinfo{pages}{3373}.
\newblock


\bibitem[\protect\citeauthoryear{Service}{Service}{[n.\,d.]}]%
        {Common_Platform}
\bibfield{author}{\bibinfo{person}{HM~Courts \&~Tribunals Service}.}
  \bibinfo{year}{[n.\,d.]}\natexlab{}.
\newblock \bibinfo{title}{HMCTS services: Common Platform}.
\newblock
\newblock
\urldef\tempurl%
\url{https://www.gov.uk/guidance/hmcts-services-common-platform#looking-ahead-in-2021}
\showURL{%
\tempurl}
\newblock
\shownote{Accessed: 2022}.


\bibitem[\protect\citeauthoryear{Service}{Service}{2021a}]%
        {prisoncategorisationpolicy2021}
\bibfield{author}{\bibinfo{person}{HM~Prison \&~Probation Service}.}
  \bibinfo{year}{2021}\natexlab{a}.
\newblock \bibinfo{title}{Security Categorisation Policy Framework}.
\newblock
\newblock
\urldef\tempurl%
\url{https://assets.publishing.service.gov.uk/government/uploads/system/uploads/attachment_data/file/1011502/security-categorisation-pf.pdf}
\showURL{%
\tempurl}


\bibitem[\protect\citeauthoryear{Service}{Service}{2021b}]%
        {metpolice2021}
\bibfield{author}{\bibinfo{person}{Metropolitan~Police Service}.}
  \bibinfo{year}{2021}\natexlab{b}.
\newblock \bibinfo{title}{Metropolitan Police Service — Written evidence
  (NTL0031)}.
\newblock
\newblock
\urldef\tempurl%
\url{https://committees.parliament.uk/writtenevidence/38736/pdf/}
\showURL{%
\tempurl}


\bibitem[\protect\citeauthoryear{Service}{Service}{2015}]%
        {OASys2015}
\bibfield{author}{\bibinfo{person}{National Offender~Management Service}.}
  \bibinfo{year}{2015}\natexlab{}.
\newblock \bibinfo{title}{A compendium of research and analysis on the Offender
  Assessment System (OASys)}.
\newblock
\newblock
\urldef\tempurl%
\url{https://assets.publishing.service.gov.uk/government/uploads/system/uploads/attachment_data/file/449357/research-analysis-offender-assessment-system.pdf}
\showURL{%
\tempurl}


\bibitem[\protect\citeauthoryear{Service}{Service}{2017}]%
        {OASys2017}
\bibfield{author}{\bibinfo{person}{National Offender~Management Service}.}
  \bibinfo{year}{2017}\natexlab{}.
\newblock \bibinfo{title}{Determining Pre Sentence Reports - Sentencing within
  the new framework. PI 04/2016}.
\newblock
\newblock
\urldef\tempurl%
\url{https://www.gov.uk/government/publications/determining-pre-sentence-reports-pi-042016}
\showURL{%
\tempurl}


\bibitem[\protect\citeauthoryear{Service}{Service}{2021c}]%
        {CPS}
\bibfield{author}{\bibinfo{person}{The Crown~Prosecution Service}.}
  \bibinfo{year}{2021}\natexlab{c}.
\newblock \bibinfo{title}{Crown Prosecution Service — Written evidence
  (NTL0018)}.
\newblock
\newblock
\urldef\tempurl%
\url{https://committees.parliament.uk/writtenevidence/38677/pdf/}
\showURL{%
\tempurl}


\bibitem[\protect\citeauthoryear{Simons, Adams~Bhatti, and Weller}{Simons
  et~al\mbox{.}}{2021}]%
        {Joshua2021}
\bibfield{author}{\bibinfo{person}{Joshua Simons}, \bibinfo{person}{Sophia
  Adams~Bhatti}, {and} \bibinfo{person}{Adrian Weller}.}
  \bibinfo{year}{2021}\natexlab{}.
\newblock \bibinfo{booktitle}{\emph{Machine Learning and the Meaning of Equal
  Treatment}}.
\newblock \bibinfo{publisher}{Association for Computing Machinery},
  \bibinfo{address}{New York, NY, USA}, \bibinfo{pages}{956–966}.
\newblock
\showISBNx{9781450384735}
\urldef\tempurl%
\url{https://doi.org/10.1145/3461702.3462556}
\showURL{%
\tempurl}


\bibitem[\protect\citeauthoryear{Society}{Society}{2018}]%
        {Law_Society}
\bibfield{author}{\bibinfo{person}{The~Law Society}.}
  \bibinfo{year}{2018}\natexlab{}.
\newblock \bibinfo{booktitle}{\emph{Algorithms in the Criminal Justice
  System}}.
\newblock \bibinfo{type}{{T}echnical {R}eport}. \bibinfo{institution}{The Law
  Society}.
\newblock
\urldef\tempurl%
\url{https://www.lawsociety.org.uk/en/topics/research/algorithm-use-in-the-criminal-justice-system-report}
\showURL{%
\tempurl}


\bibitem[\protect\citeauthoryear{Society}{Society}{2021}]%
        {royalsociety}
\bibfield{author}{\bibinfo{person}{The~Royal Society}.}
  \bibinfo{year}{2021}\natexlab{}.
\newblock \bibinfo{title}{Privacy Enhancing Technologies}.
\newblock
\newblock
\urldef\tempurl%
\url{https://royalsociety.org/topics-policy/projects/privacy-enhancing-technologies/}
\showURL{%
\tempurl}


\bibitem[\protect\citeauthoryear{SpaceTimeAI}{SpaceTimeAI}{[n.\,d.]}]%
        {iHotSpot}
\bibfield{author}{\bibinfo{person}{SpaceTimeAI}.}
  \bibinfo{year}{[n.\,d.]}\natexlab{}.
\newblock \bibinfo{title}{iHotSpot}.
\newblock
\newblock
\urldef\tempurl%
\url{http://spacetimeai.com/iHotSpot.html}
\showURL{%
\tempurl}
\newblock
\shownote{Accessed: 2021}.


\bibitem[\protect\citeauthoryear{Sprick}{Sprick}{2019}]%
        {sprick2019predictive}
\bibfield{author}{\bibinfo{person}{Daniel Sprick}.}
  \bibinfo{year}{2019}\natexlab{}.
\newblock \showarticletitle{Predictive Policing in China: An Authoritarian
  Dream of Public Security}.
\newblock \bibinfo{journal}{\emph{Navei{\~n} Reet: Nordic Journal of Law and
  Social Research (NNJLSR) No}}  \bibinfo{volume}{9} (\bibinfo{year}{2019}).
\newblock


\bibitem[\protect\citeauthoryear{Steria}{Steria}{2019}]%
        {STORM_Merseyside_Police}
\bibfield{author}{\bibinfo{person}{Sopra Steria}.}
  \bibinfo{year}{2019}\natexlab{}.
\newblock \bibinfo{title}{Merseyside Police sign contract for STORM Command and
  Control system}.
\newblock
\newblock
\urldef\tempurl%
\url{https://www.soprasteria.co.uk/thinking/blogs/details/merseyside-police-sign-contract-for-storm-command-and-control-system}
\showURL{%
\tempurl}


\bibitem[\protect\citeauthoryear{Szegedy, Zaremba, Sutskever, Bruna, Erhan,
  Goodfellow, and Fergus}{Szegedy et~al\mbox{.}}{2014}]%
        {szegedy2014intriguing}
\bibfield{author}{\bibinfo{person}{Christian Szegedy},
  \bibinfo{person}{Wojciech Zaremba}, \bibinfo{person}{Ilya Sutskever},
  \bibinfo{person}{Joan Bruna}, \bibinfo{person}{Dumitru Erhan},
  \bibinfo{person}{Ian Goodfellow}, {and} \bibinfo{person}{Rob Fergus}.}
  \bibinfo{year}{2014}\natexlab{}.
\newblock \bibinfo{title}{Intriguing properties of neural networks}.
\newblock
\newblock
\showeprint[arxiv]{1312.6199}~[cs.CV]


\bibitem[\protect\citeauthoryear{Technology}{Technology}{[n.\,d.]}]%
        {Niche}
\bibfield{author}{\bibinfo{person}{Niche Technology}.}
  \bibinfo{year}{[n.\,d.]}\natexlab{}.
\newblock \bibinfo{title}{Niche RMS Customer Profiles}.
\newblock
\newblock
\urldef\tempurl%
\url{https://nicherms.com/region/uk/}
\showURL{%
\tempurl}
\newblock
\shownote{Accessed: 2021}.


\bibitem[\protect\citeauthoryear{the Comptroller and Genera}{the Comptroller
  and Genera}{2016}]%
        {Audit2016}
\bibfield{author}{\bibinfo{person}{the Comptroller} {and}
  \bibinfo{person}{Auditor Genera}.} \bibinfo{year}{2016}\natexlab{}.
\newblock \bibinfo{booktitle}{\emph{Efficiency in the criminal justice
  system}}.
\newblock \bibinfo{type}{{T}echnical {R}eport}. \bibinfo{institution}{National
  Audit Office}.
\newblock
\urldef\tempurl%
\url{https://www.nao.org.uk/wp-content/uploads/2016/03/Efficiency-in-the-criminal-justice-system.pdf}
\showURL{%
\tempurl}


\bibitem[\protect\citeauthoryear{Trust}{Trust}{2021a}]%
        {prisonreformtrust2021}
\bibfield{author}{\bibinfo{person}{Prison~Reform Trust}.}
  \bibinfo{year}{2021}\natexlab{a}.
\newblock \bibinfo{title}{New technologies and the application of the law:
  Written evidence}.
\newblock
\newblock
\urldef\tempurl%
\url{https://committees.parliament.uk/writtenevidence/38265/pdf/}
\showURL{%
\tempurl}


\bibitem[\protect\citeauthoryear{Trust}{Trust}{2018}]%
        {Prison_Reform2018}
\bibfield{author}{\bibinfo{person}{The Prison~Reform Trust}.}
  \bibinfo{year}{2018}\natexlab{}.
\newblock \bibinfo{title}{Offender Management and Sentence Plan}.
\newblock
\newblock
\urldef\tempurl%
\url{http://www.prisonreformtrust.org.uk/ForPrisonersFamilies/PrisonerInformationPages/OffenderManagementandsentenceplanning}
\showURL{%
\tempurl}


\bibitem[\protect\citeauthoryear{Trust}{Trust}{2021b}]%
        {Prison_Reform}
\bibfield{author}{\bibinfo{person}{The Prison~Reform Trust}.}
  \bibinfo{year}{2021}\natexlab{b}.
\newblock \bibinfo{title}{Prison Reform Trust — Written evidence (NTL0004)}.
\newblock
\newblock
\urldef\tempurl%
\url{https://committees.parliament.uk/writtenevidence/38265/pdf/}
\showURL{%
\tempurl}


\bibitem[\protect\citeauthoryear{(UK)}{(UK)}{2021}]%
        {Home_Office2021}
\bibfield{author}{\bibinfo{person}{The Home~Office (UK)}.}
  \bibinfo{year}{2021}\natexlab{}.
\newblock \bibinfo{title}{Home Office — Written evidence (NTL0055)}.
\newblock
\newblock
\urldef\tempurl%
\url{https://committees.parliament.uk/writtenevidence/40975/pdf/}
\showURL{%
\tempurl}


\bibitem[\protect\citeauthoryear{Vestby and Vestby}{Vestby and Vestby}{2021}]%
        {machinelearningvestby}
\bibfield{author}{\bibinfo{person}{Annette Vestby} {and} \bibinfo{person}{Jonas
  Vestby}.} \bibinfo{year}{2021}\natexlab{}.
\newblock \showarticletitle{Machine Learning and the Police: Asking the Right
  Questions}.
\newblock \bibinfo{journal}{\emph{Policing: A Journal of Policy and Practice}}
  \bibinfo{volume}{15} (\bibinfo{year}{2021}), \bibinfo{pages}{44}.
\newblock
Issue 1.


\bibitem[\protect\citeauthoryear{Watch}{Watch}{2019}]%
        {bigbrother2019}
\bibfield{author}{\bibinfo{person}{Big~Brother Watch}.}
  \bibinfo{year}{2019}\natexlab{}.
\newblock \bibinfo{title}{Big Brother Watch’s written evidence on algorithms
  in the justice system for the Law Society’s Technology and the Law Policy
  Commission}.
\newblock
\newblock
\urldef\tempurl%
\url{https://bigbrotherwatch.org.uk/wp-content/uploads/2019/02/Big-Brother-Watch-written-evidence-on-algorithms-in-the-justice-system-for-the-Law-Societys-Technology-and-the-Law-Policy-Commission-Feb-2019.pdf}
\showURL{%
\tempurl}


\bibitem[\protect\citeauthoryear{Weisburd, Greenspan, Mastrofski, and
  Willis}{Weisburd et~al\mbox{.}}{2008}]%
        {weisburd2008compstat}
\bibfield{author}{\bibinfo{person}{David Weisburd}, \bibinfo{person}{Rosann
  Greenspan}, \bibinfo{person}{Stephen Mastrofski}, {and} \bibinfo{person}{J
  Willis}.} \bibinfo{year}{2008}\natexlab{}.
\newblock \bibinfo{title}{Compstat and organizational change: A national
  assessment}.
\newblock
\newblock


\bibitem[\protect\citeauthoryear{Weller}{Weller}{2019}]%
        {weller2019transparency}
\bibfield{author}{\bibinfo{person}{Adrian Weller}.}
  \bibinfo{year}{2019}\natexlab{}.
\newblock \bibinfo{title}{Transparency: motivations and challenges, in
  ‘Explainable AI: Interpreting, Explaining and Visualizing Deep
  Learning’}.
\newblock
\newblock


\end{thebibliography}

\appendix
\beginsupplement

\end{document}